\documentclass[aps,preprint]{revtex4}

\usepackage[dvips]{graphicx}

\newcommand{\ch}{{\cal H}}
\newcommand{\ce}{{\cal E}}

\begin{document}

\date{\today}

\title{Quantum states talk via the environment}
\author{Hichem Eleuch$^{1,2}$\footnote{email: hichemeleuch@yahoo.fr} and 
Ingrid Rotter$^{3}$\footnote{email: rotter@pks.mpg.de}}

\address{
$^1$Department of Physics, McGill University, Montreal, Canada H3A 2T8}
\address{
$^{2}$ Department of Physics, Universit\'{e} de Montr\'{e}al, Montreal, QC,  H3T 1J4, Canada
}
\address{
$^3$Max Planck Institute for the Physics of Complex Systems,
D-01187 Dresden, Germany }

\begin{abstract}
  
The states of an open quantum system interact (``talk'') with one another 
via the extended environment into which the localized system is embedded. 
This interaction is mediated by the source term of the Schr\"odinger 
equation which describes the coupling between system
and environment. The source term is nonlinear and causes width bifurcation
and, respectively, level repulsion. It is
strong only in the neighborhood of singular (exceptional) points. 
We provide  typical results for the phase rigidity and the mixing 
of the biorthogonal eigenfunctions of the Hamiltonian.
A completely unexpected result is that the phase rigidity approaches a value
near to one (characteristic of orthogonal 
eigenfunctions) when width bifurcation (or level repulsion) becomes
maximum. This behavior of the phase rigidity 
is caused exclusively by the nonlinearity of the source
term of the Schr\"odinger equation. The eigenfunctions remain mixed in
the set of original wavefunctions also under these critical conditions.
Eventually, a dynamical phase transition occurs. This process is 
irreversible. It allows, among others, a physical interpretation of the 
well-known resonance trapping phenomenon. 
Our results for the eigenvalues and eigenfunctions of a non-Hermitian 
Hamiltonian are supported by experimental results obtained in
different systems. The relation of our results for open quantum
systems under critical conditions to those in optics and photonics with
PT-symmetry breaking is considered. As a result, the balance between 
gain and loss is a very interesting general phenomenon that may occur
in many different systems.

\end{abstract}

\pacs{\bf }
\maketitle

\section{Introduction}
\label{intr}

Physical systems with PT-symmetry and its breaking 
are studied recently in very many papers, 
see e.g. the review {\it Making sense of non-Hermitian 
Hamiltonians} \cite{bender}. Using the equivalence
between the one-particle Schr\"odinger equation and the optical wave
equation \cite{equivalence}, it was possible to prove the theoretical 
results in optical devices \cite{ptsymmexp}. Further studies provided
interesting applications in different PT-symmetric systems. This
caused some new trends in recent PT-symmetry studies:
{\it Although PT-symmetric systems were originally explored at a
highly mathematical level, it is now understood that one can interpret
PT-symmetric systems simply as nonisolated physical systems having a
balanced loss and gain}  \cite{maria}.

In literature, open quantum systems are described  
usually by means of a non-Hermitian Hamiltonian,
e.g. \cite{ro91,moiseyev,auerbach} and also \cite{top}. In these papers,
PT-symmetry does not play a special role. Nevertheless, physical processes 
similar to PT-symmetry breaking are observed experimentally. They are 
called mostly  dynamical phase transitions \cite{pastawski,top},
sometimes also superradiance  \cite{auerbach}. Most convincing
experimental results are obtained
in studies on mesoscopic systems, see the review \cite{ropp}.  

In calculations for concrete systems, the non-Hermiticity of the 
Hamilton operator is introduced mostly by 
adding a non-Hermitian perturbation term to the 
Hermitian Hamiltonian that describes the main features of the system,
e.g.  \cite{atabek,jogle}.
Due to dynamical phase transitions appearing in open quantum systems,
the definition of this Hermitian Hamiltonian is
however not unique. For example, it is basically a shell-model 
Hamiltonian when light nuclei 
are considered \cite{nobel}  while it is a completely other
Hamiltonian in heavy nuclei \cite{feshbach}. In the last case, the 
individual states are described usually by statistical assumptions 
(mostly according to random matrix theory). 

Generally, open quantum systems consist of some localized microscopic 
region that is embedded into an infinitely large environment of scattering
wavefunctions. Due to this embedding, the states of the localized 
region can decay into the environment giving them a  finite lifetime. 
The environment can be changed,  it can however never be deleted. This
can be seen immediately from the finite 
lifetime of, e.g., nuclear states which is determined exclusively 
(without participation of any external observer) by the 
wavefunctions of the individual states and their coupling to the
environment. The lifetime of the nuclear states
can be used therefore for radioactive geologic age determination.

According to this statement, the
whole function space of the open quantum system consists of two parts: 
one part is the function space of 
discrete states of the localized part while the other part is the
function space of continuous scattering wavefunctions.
A very successful method used to describe the properties of an open 
quantum system, is therefore a formalism with
two projection operators  each of which is related to one of the two
parts of the function space.  This method developed in nuclear
physics many years ago \cite{feshbach}, has been applied also to the
description of other systems, see e.g. the review \cite{top}.  
The Hamilton operator of the whole system is Hermitian while those 
two Hamiltonians that describe the 
properties of either subsystem, are non-Hermitian.

The question arises  what is the relation between PT-symmetry
breaking and a dynamical phase transition appearing in an open quantum
system. In both cases, the Hamiltonian of the system is non-Hermitian.
In the first case, the eigenvalues are real and become complex under
critical conditions. In the second case, critical conditions arise 
at and near to mathematical singularities which cause essential changes 
in the eigenfunctions of the Hamiltonian. These changes may occur
because the eigenfunctions are biorthogonal and the 
states of the system may talk via the environment.

It is the aim of the present paper to study  differences and 
similarities between the physical results obtained for non-Hermitian
quantum systems under critical conditions and that, on the one hand, 
in the case of PT-symmetry breaking  and, on the other hand, 
in the case of dynamical phase transitions. 
In contrast to most other models for the description of open quantum
systems, we do not start from a specific 
Hermitian Hamiltonian. Instead, we consider a 
Hamiltonian that is completely non-Hermitian from the very beginning;  
determine its eigenvalues and eigenfunctions;
investigate nonlinear effects (that cause  irreversibility);
and trace  dynamical phase transitions.  
The formalism is described in \cite{top,nearby1,nearby2}.
In the present paper,
we are interested, above all, in the fact that the states of
an open quantum system can talk to one another via
the environment into which the system is embedded.
By this, the wavefunctions become mixed and their phases cease to
be rigid. This is a new feature 
characteristic of any open quantum  system. It is incompatible 
with the basic assumptions of Hermitian quantum physics.
 
In section \ref{ham}, we provide  expressions for the
eigenvalues and eigenfunctions of a
non-Hermitian $2 \times 2$ operator $\ch^{(2)}$
and point shortly to their  specific features by which they 
differ from the eigenvalues and eigenfunctions
of a Hermitian operator. The eigenfunctions of a non-Hermitian
operator are biorthogonal. Most important are the singular points,
called usually {\it  exceptional points (EPs)}, at which the
biorthogonality of the eigenfunctions plays an important role. 
In the following
section \ref{eig}, we consider the eigenfunctions of  $\ch^{(2)}$
in the very neighborhood of EPs. Above all, we consider the
phase rigidity of the eigenfunctions which is a quantitative measure
of the difference  between orthogonal and biorthogonal eigenfunctions.
Under the influence of EPs, it is reduced and vanishes at an EP.
Here, the eigenfunctions are mixed in the set of basic wavefunctions. 
Some numerical results for the phase rigidity and for the
mixing of the wavefunctions are  shown and discussed in section 
\ref{num} for natural systems the states of which 
can decay, as well as for systems with loss and gain. Most
astonishing and unexpected result is that the eigenfunctions are 
almost orthogonal at the critical point of maximum width 
bifurcation or maximum level repulsion. We discuss these results 
in section \ref{disc}. In the following section \ref{trap}, their 
relation to the phenomenon of resonance trapping is considered. 
Some concluding remarks can be found in section  \ref{concl}. 
According to our results, balanced loss and gain appears not 
only in PT-symmetric systems in optics and photonics, 
but is a more general phenomenon that can be seen also in other
systems. It is therefore of great value for applications.

\section{Eigenvalues and eigenfunctions of a non-Hermitian Hamilton operator}
\label{ham}

The calculation of the eigenvalues $\ce_i$ and
eigenfunctions $\Phi_i$ of a non-Hermitian Hamiltonian $\ch$
hits upon some mathematically non-trivial problems due to the
existence of  singular points  in the continuum. At these 
points, two eigenvalues coalesce and are supplemented by 
an associated vector defined by the
Jordan chain relations \cite{gurosa}. The
two corresponding eigenfunctions differ
from one another only by a phase \cite{top,comment1}. 
The geometric phase of these points
differs from the Berry phase of a diabolic point
by  a factor 2. These singular points,
well-known in mathematics \cite{kato}, are called mostly
exceptional points (EPs).
Their meaning for the dynamics of open quantum systems 
and the behavior of the two eigenfunctions at and near to an EP is
studied only recently.

Let us consider, as an example, the symmetric $2\times 2$ matrix 
\begin{eqnarray}
{\cal H}^{(2)} = 
\left( \begin{array}{cc}
\varepsilon_{1} \equiv e_1 + \frac{i}{2} \gamma_1  & ~~~~\omega_{12}   \\
\omega_{21} & ~~~~\varepsilon_{2} \equiv e_2 + \frac{i}{2} \gamma_2   \\
\end{array} \right) 
\label{form1}
\end{eqnarray}
with $\gamma_{i=1,2} \le 0$  or with $\gamma_1 \le 0$ and 
 $\gamma_2 \ge 0$ \cite{comment2}. The diagonal elements of (\ref{form1}) 
are the two complex eigenvalues 
$ \varepsilon_{i}~(i=1,2)$ of the non-Hermitian operator ${\cal H}_0^{(2)}$. 
That means, the $e_i$ and  $\gamma_i$ denote the 
energies and widths, respectively, of the two states when
$\omega_{ij} =0$. The $\omega_{12}=\omega_{21}\equiv \omega$ stand for
the coupling matrix elements of the two states via the common
environment which are, generally,
complex \cite{top,nearby1}. The selfenergy of the
states is assumed to be included into the $\varepsilon_i$.
The Hamiltonian $\ch^{(2)}$ allows us to consider the properties of
the system near to and at an EP because here the distance between
the two states, that coalesce at the EP, relative to one another is
much smaller than that relative to the other states of the system.

The eigenvalues of $\ch^{(2)}$ are
\begin{eqnarray} 
\ce_{i,j} ~\equiv ~ E_{i,j} + \frac{i}{2} \Gamma_{i,j} ~ = ~ 
 \frac{\varepsilon_1 + \varepsilon_2}{2} \pm Z ~; \quad \quad
Z \equiv \frac{1}{2} \sqrt{(\varepsilon_1 - \varepsilon_2)^2 + 4 \omega^2}
\label{int6}
\end{eqnarray}
where  $E_{i,j}$ and $\Gamma_{i,j}$ stand for the
energy and width, respectively, of the eigenstates $i$ 
and $j$ \cite{comment2}.
The parametrical variation of the eigenvalues at the EP does not 
follow Fermi's golden rule. Instead, the resonance states  
repel each other in energy  according to  Re$(Z) $
while the widths bifurcate according to  Im$(Z)  $.
The transition from level repulsion to width bifurcation is studied 
numerically in e.g. \cite{nearby1,elro3}. The two states cross 
when $Z=0$. Here, the two eigenvalues  coalesce, $\ce_{1}=\ce_{2}$;
are supplemented by an associated vector \cite{gurosa}; 
and the crossing point is 
an EP in agreement with the definition of Kato \cite{kato}.

Further, the eigenfunctions of a non-Hermitian $\ch$ 
must fulfill the conditions $\ch|\Phi_i\rangle =
  {\ce}_i|\Phi_i\rangle$ and $\langle \Psi_i|\ch = {\ce}_i \langle
  \Psi_i|$ where  $\ce_i$ is an eigenvalue of $\ch$ and the vectors 
$ |\Phi_i\rangle$ and $\langle \Psi_i|$ denote its right and left
eigenfunctions, respectively. When $\ch$ is a Hermitian operator, the
$\ce_i$ are real, and we arrive at the
 well-known relation $\langle \Psi_i| =  \langle \Phi_i|$. In this case, 
the eigenfunctions can be normalized by using the expression 
$\langle \Phi_i|\Phi_j\rangle$. 
For the symmetric non-Hermitian Hamiltonian $\ch^{(2)}$, however, we
have  $\langle \Psi_i| =  \langle \Phi_i^*|$. This means, that the
eigenfunctions are biorthogonal and have to be normalized by means of 
$\langle \Phi_i^*|\Phi_j\rangle$. This is, generally, 
a {\it complex} value, in contrast to the real value 
 $\langle \Phi_i|\Phi_j\rangle$ of the Hermitian case. To smoothly
describe the transition from a closed system with discrete states, to
a weakly open one with narrow resonance states, we normalize the 
$\Phi_i$  according to 
\begin{eqnarray}
\langle \Phi_i^*|\Phi_j\rangle = \delta_{ij} 
\label{int3}
\end{eqnarray}
(for details see sections 2.2 and 2.3 of \cite{top}). It follows  
\begin{eqnarray}
 \langle\Phi_i|\Phi_i\rangle & = & 
{\rm Re}~(\langle\Phi_i|\Phi_i\rangle) ~; \quad
A_i \equiv \langle\Phi_i|\Phi_i\rangle \ge 1
\label{int4} 
\end{eqnarray}
and 
\begin{eqnarray}
\langle\Phi_i|\Phi_{j\ne i}\rangle & = &
i ~{\rm Im}~(\langle\Phi_i|\Phi_{j \ne i}\rangle) =
-\langle\Phi_{j \ne i}|\Phi_i\rangle 
\nonumber  \\
&& |B_i^j|  \equiv 
|\langle \Phi_i | \Phi_{j \ne i}| ~\ge ~0  \; .
\label{int5}
\end{eqnarray}
At an EP $A_i \to \infty$ and $|B_i^j| \to \infty$.
The $\Phi_i$ contain (like the $\ce_i$)  global features that are 
caused by many-body forces  induced by the coupling
$\omega_{ik}$ of the states $i$ and $k\ne i$ via the environment
(which has an infinite number of degrees of freedom).
The eigenvalues $\ce_i$ and eigenfunctions $\Phi_i$ contain 
moreover the self-energy contributions of the states $i$
due to their coupling to the environment. 

The Schr\"odinger equation with the non-Hermitian operator 
${\cal H}^{(2)}$ is equivalent to a Schr\"odinger equation with 
${\cal H}_0^{(2)}$ and source term 
\begin{eqnarray}
\label{form1a}
({\cal H}_0^{(2)} - \varepsilon_i) ~| \Phi_i \rangle  = -
\left(
\begin{array}{cc}
0 & \omega_{ij} \\
\omega_{ji} & 0
\end{array} \right) |\Phi_j \rangle \equiv W  |\Phi_j \rangle\; . 
\end{eqnarray}
Due to the source term, two states are coupled via the 
common environment of scattering wavefunctions into which the system 
is embedded,  $\omega_{ij}=\omega_{ji}\equiv\omega$.
The Schr\"odinger equation (\ref{form1a}) with source term can be
rewritten in the following manner \cite{ro01},
\begin{eqnarray}
\label{form2a}
({\cal H}_0^{(2)}  - \varepsilon_i) ~| \Phi_i \rangle  = 
\sum_{k=1,2} \langle
\Phi_k|W|\Phi_i\rangle \sum_{m=1,2} \langle \Phi_k |\Phi_m\rangle 
|\Phi_m\rangle \; . 
\end{eqnarray}
According to the biorthogonality  relations
(\ref{int4}) and (\ref{int5}) of the eigenfunctions of ${\cal H}^{(2)}$,  
(\ref{form2a}) is a nonlinear equation.  
Most important part of the nonlinear contributions is contained in 
\begin{eqnarray}
\label{form3a}
({\cal H}_0^{(2)}  - \varepsilon_n) ~| \Phi_n \rangle =
\langle \Phi_n|W|\Phi_n\rangle ~|\Phi_n|^2 ~|\Phi_n\rangle \; .  
\end{eqnarray}
The nonlinear source term vanishes far from an EP where
$\langle \Phi_k|\Phi_{k }\rangle    \to  1 $ and
$\langle \Phi_k|\Phi_{l\ne k }\rangle = - 
\langle \Phi_{l \ne k  }|\Phi_{k}\rangle \to  0 $
as follows from the normalization (\ref{int3}).
Thus, the Schr\"odinger equation with source term is 
(almost) linear far from an EP, as usually assumed. It is however 
nonlinear in the neighborhood of an EP.

It is meaningful to represent
the  eigenfunctions $\Phi_i$ of ${\cal H}^{(2)}$  in the
set of basic wavefunctions $\Phi_i^0$ of ${\cal H}_0^{(2)}$
\begin{eqnarray}
\Phi_i=\sum_{j=1}^N b_{ij} \Phi_j^0 ~~ ;
\quad \quad b_{ij} = |b_{ij}| e^{i\theta_{ij}}
\; .
\label{int20}
\end{eqnarray}
Also the $b_{ij}$ are normalized  according to the biorthogonality
relations  of the wavefunctions $\{\Phi_i\}$. The angle $\theta_{ij}$
can be determined from
${\rm tg}(\theta_{ij}) = {\rm Im}(b_{ij}) / {\rm Re}(b_{ij})$ .

\section{Eigenfunctions of a non-Hermitian operator near to an exceptional point}
\label{eig}

An EP is a point in the continuum and therefore of measure zero. It
can be identified only due to its influence on the eigenstates of the
localized system in a certain finite parameter range around the EP.
This influence is visible, indeed, as the results of 
many calculations have shown (see
e.g. \cite{top,nearby1,nearby2}). One of the consequences is that more
than two states cannot cross in one point since  every state near to
an EP will interact with states that are modified by the EP (and not
with the original states). Under this condition, the
ranges of influence of different EPs overlap, meaning that some
clustering of EPs occurs \cite{nearby2,cluster}.

At the EP, the eigenfunctions $\Phi_i^{\rm cr}$ of ${\cal H}^{(2)}$
of the two crossing states differ from one another only by a phase, 
\begin{eqnarray}
\Phi_1^{\rm cr} \to ~\pm ~i~\Phi_2^{\rm cr} \; ;
\quad \qquad \Phi_2^{\rm cr} \to
~\mp ~i~\Phi_1^{\rm cr}   
\label{eif5}
\end{eqnarray}  
according to analytical  as well as numerical and experimental
studies \cite{comment}.
That means, the wavefunction $\Phi_1$ of the state $1$ jumps
at the EP to   ~$\pm\, i\, \Phi_2$. 

The biorthogonality of the eigenfunctions $\Phi_k$ of the non-Hermitian
operator $\ch^{(2)}$ is characterized quantitatively by the ratio 
\begin{eqnarray}
r_k ~\equiv ~\frac{\langle \Phi_k^* | \Phi_k \rangle}{\langle \Phi_k 
| \Phi_k \rangle} ~= ~A_k^{-1} \; . 
\label{eif11}
\end{eqnarray}
We call  $r_k$, defined by (\ref{eif11}), the {\it phase
rigidity}  of the eigenfunction $\Phi_k$. 
Generally $1 ~\ge ~r_k ~\ge ~0 $. For decaying states which are well
separated  from other decaying states, it holds $r_k \approx 1$.
This result corresponds to the fact that Hermitian quantum physics is 
a good approximation at low level density. The situation changes however 
completely when an EP is approached\,:
\begin{verse}
(i) When  two levels
 are distant from one another,  their eigenfunctions
 are (almost) orthogonal,  
$\langle \Phi_k^* | \Phi_k \rangle   \approx
\langle \Phi_k | \Phi_k \rangle  \equiv A_k \approx 1 $.\\
(ii) When  two levels cross at the EP, their eigenfunctions are linearly
dependent according to (\ref{eif5}) and 
$\langle \Phi_k | \Phi_k \rangle \equiv A_k \to \infty $.\\
\end{verse}
These two relations show that the phases of the two eigenfunctions
relative to one another change dramatically 
when the crossing point (EP) is approached. The  non-rigidity of the
phases of the eigenfunctions of $\ch^{(2)}$ expressed by $r_k < 1$, 
follows, of course, directly from the fact that 
$\langle\Phi_k^*|\Phi_k\rangle$ is a complex number (in difference to 
the norm $\langle\Phi_k|\Phi_k\rangle$ which is a real number) 
such that the normalization condition
(\ref{int3}) can be fulfilled only by the additional postulation 
Im$\langle\Phi_k^*|\Phi_k\rangle =0$ (what corresponds to a rotation). 

Mathematically, $r_k<1$ causes nonlinear effects in 
quantum systems in a natural manner according to (\ref{form3a}).
When $r_k<1$, an analytical expression for the eigenfunctions as 
function of a certain control parameter  can, generally, not be
obtained. 

The variation of the rigidity, $0< r_k< 1$, of the phases of the 
eigenfunctions of $\ch^{(2)}$ in the neighborhood of EPs is the most important
difference between the  non-Hermitian quantum physics and the Hermitian
one. It expresses the fact that two nearby states of an open quantum
system can strongly interact with one another via the environment\,:
\begin{verse}
(i) Two orthogonal eigenstates of a Hermitian operator can be mixed only by 
means of a direct interaction $V^{\rm res}$ between the two states because of 
the orthogonality relation $\langle \Psi_k|\Psi_{i \ne k} \rangle 
\equiv |B_k^i|=0$. The direct interaction  $V^{\rm res}$ causes   
$\langle \Psi_k|V^{\rm res}|\Psi_{i \ne k} \rangle \ne 0$.\\
(ii) Two biorthogonal eigenstates of a non-Hermitian operator can,
additio\-nally, mix via the common environment of scattering
wavefunctions. It is
$\langle \Phi_k|\Phi_{i \ne k} \rangle \equiv |B_k^i| \ge 0$
because of the biorthogonality condition
$\langle \Phi_k^*|\Phi_{i \ne k} \rangle = 0$, Eq. (\ref{int3}).  \\
\end{verse}
According to the relations (\ref{int5}), the mixing of the states via
the environment is large only in 
the neighborhood of EPs. That means, two quantum states of a localized 
system talk via the environment at all parameter values, indeed, however  
in a substantial manner only when they are neighbored, i.e. when they 
are near to an EP. 

We underline here that the
mixing, expressed quantitatively by (\ref{int20}), is caused by the coupling 
of {\it all} states of the localized system to the common extended environment. 
It is a feature  characteristic of open quantum systems described by a 
non-Hermitian Hamiltonian with biorthogonal eigenfunctions; and does not appear 
in Hermitian quantum physics with orthogonal eigenfunctions.

\section{Analytical and numerical results}
\label{num}

\begin{figure}[ht]
 \begin{center}
\includegraphics[width=8cm,height=13cm]{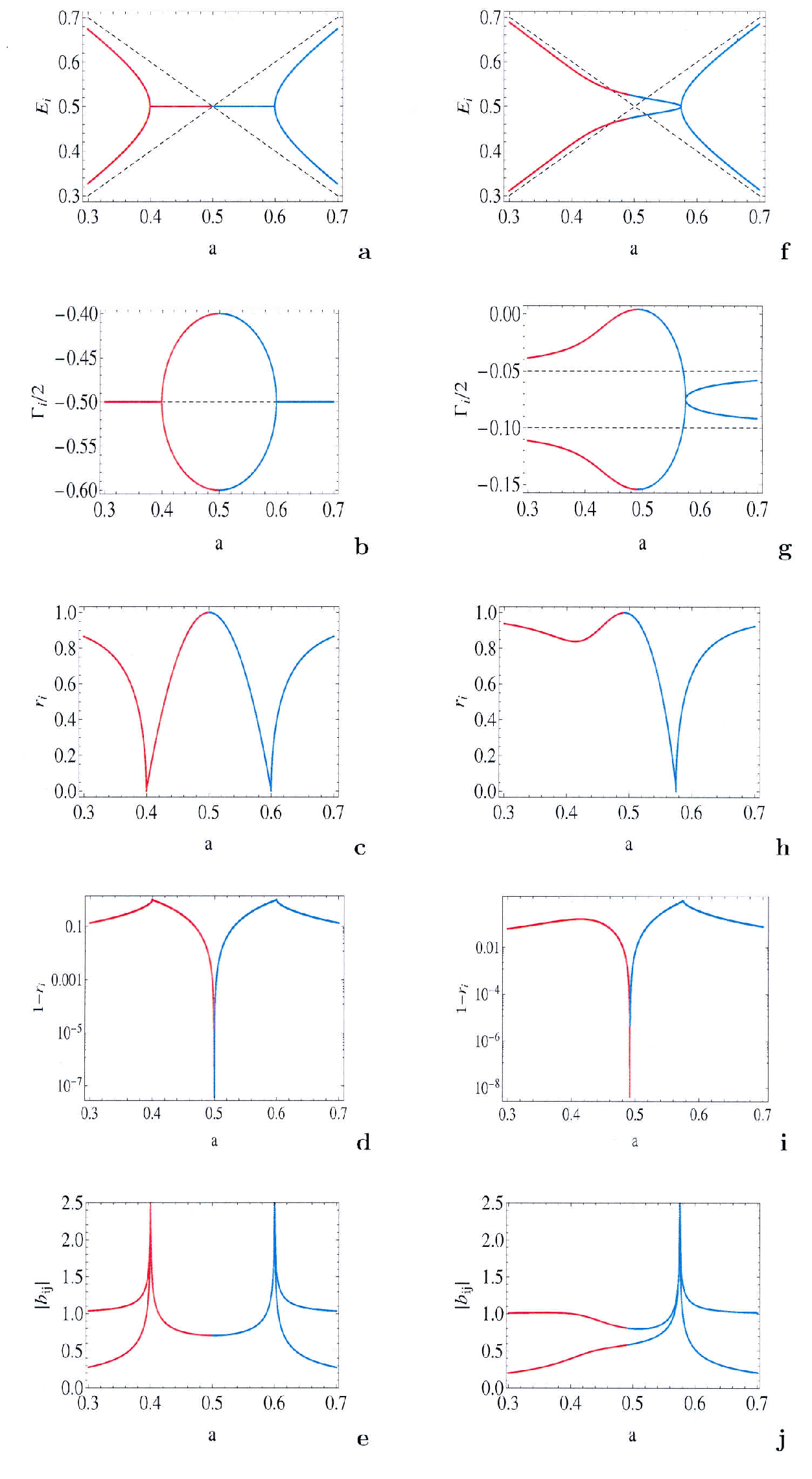}
\vspace{-.6cm}
\end{center}
\caption{\footnotesize
Energies $E_i$ (a,f);  widths $\Gamma_i/2$ (b,g);
phase rigidity $r_i$ (c,h); $1-r_i$  (d,i); and
mixing coefficients $|b_{ij}|$ of the wavefunctions 
$\Phi_i$ and $\Phi_j$  of $N=2$ states coupled to a
common channel as a function of $a$.\\
 Parameters: 
$~e_1=1-a;~e_2=a;~\gamma_1/2=\gamma_2/2=-0.5; 
~\omega=0.1\,i $ (left panel) and
$~e_1=1-a;~e_2=a; ~\gamma_1/2=-0.05; ~\gamma_2/2=-0.1;
~\omega=0.1\,(\frac{1}{4} + \frac{3}{4}\,i)$
(right panel).\\
The dashed lines in (a, b, f, g) show $e_i(a)$
and $\gamma_i(a)$, respectively.
The phase rigidity $r_i$ approaches zero at an EP, while it approaches
the value one at the point of maximum width bifurcation. The
eigenfunctions are mixed in both cases. When $r_i \to 1$, the wavefunctions are
almost orthogonal. At this point, we have changed the color of the
different trajectories in order to express the irreversibility of the parametric
evolution up to this point (see the discussion in Sect. \ref{disc}).
}
\label{fig1}
\end{figure}

\begin{figure}[ht]
\begin{center}
\includegraphics[width=8cm,height=13cm]{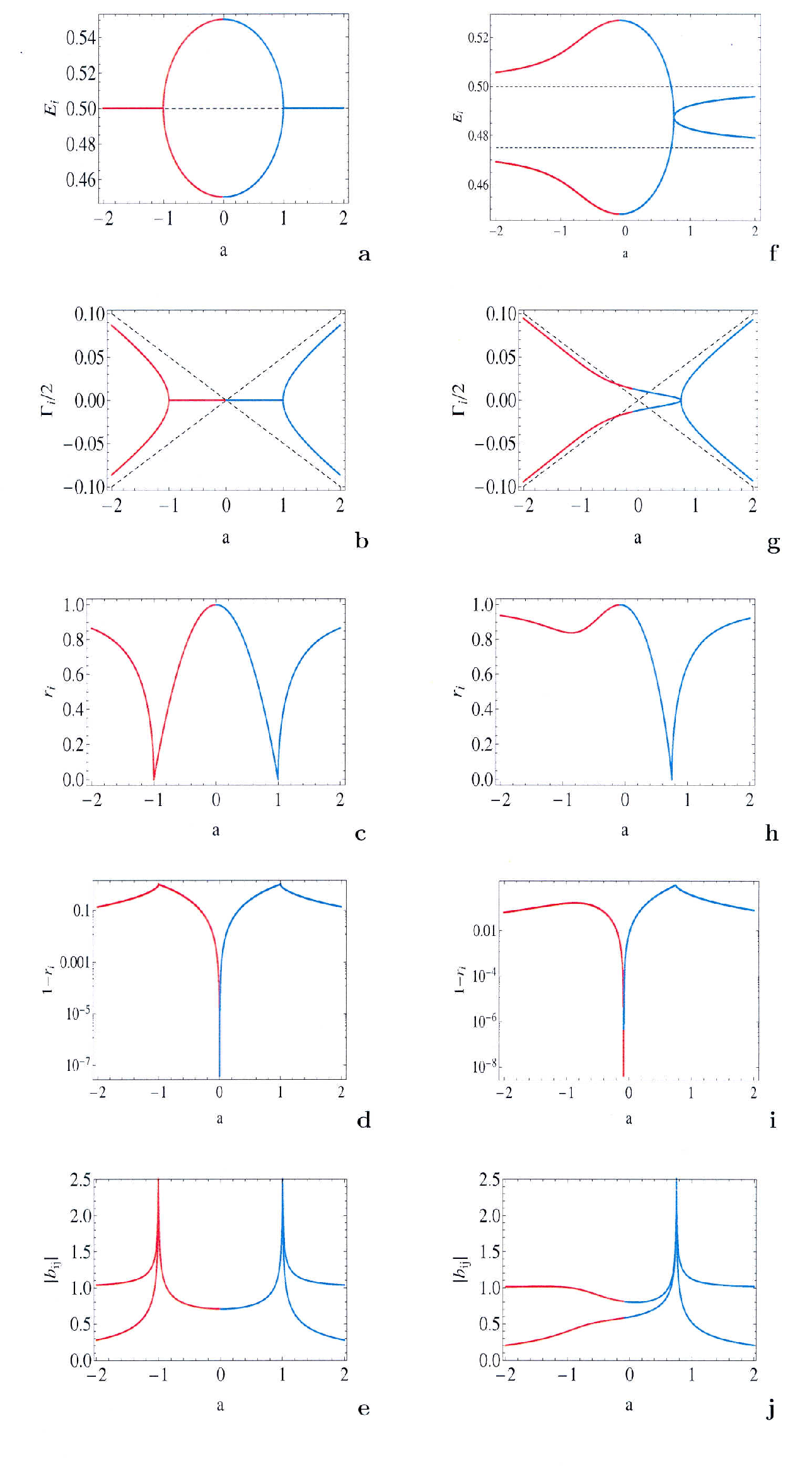}
\vspace{-.6cm}
\end{center}
\caption{\footnotesize
The same as Fig. \ref{fig1} but
$~e_1 = e_2= 0.5; ~\gamma_1 =-0.05\,a; \gamma_2 = 0.05\,a $;
~$\omega = 0.05$   (left panel);  
$~e_1 = 0.5; ~e_2 = 0.475;
 ~\gamma_1 = -0.05\, a ; ~\gamma_2 =0.05\,a$; 
~$\omega = 0.05\,(\frac{3}{4} + \frac{1}{4} \,i)$ (right panel).\\
The phase rigidity $r_i$ approaches zero at an EP, while it approaches
the value one at the point of maximum level repulsion.
The results for the phase rigidity $r_i$ and the mixing coefficients 
$|b_{ij}|$ are analog to those in Fig. \ref{fig1}.
}
\label{fig2}
\end{figure}

In order to illustrate how  the states of a quantum
system talk through the environment, we provide and discuss some results 
obtained analytically and numerically for nearby states.
We start from the $2\times 2$ Hamiltonian  (\ref{form1}) for the
description of the system. First we consider
a natural system with decaying states ($\gamma_{i=1,2} < 0$\,;
~Fig. \ref{fig1}) and secondly  a system with loss and gain  
($\gamma_1 < 0\,; ~\gamma_2>0$\,; ~Fig. \ref{fig2}). 
We show the real  and imaginary parts ($E_i$ and $\Gamma_i$, respectively)  
of the eigenvalues of $\ch^{(2)}$ as function of
a certain parameter together with the corresponding 
phase rigidity $r_i$ of the eigenfunctions $\Phi_i$ and the mixing 
coefficients $|b_{ij}|$ of the two eigenfunctions $\Phi_i$ and
$\Phi_j$ via the continuum.
We choose the  non-diagonal matrix elements $\omega_{12} = \omega_{21}
\equiv \omega$ in  (\ref{form1}) to be
complex in Figs. \ref{fig1} and \ref{fig2}, right panels, 
according to the complex coupling coefficients between system
and environment in  realistic systems \cite{top}. 
For comparison with analytical results,  the calculations in 
the left panels of  Figs. \ref{fig1} and \ref{fig2}
are performed with, respectively, imaginary and real $\omega$. In any
case, {\it the $\omega$ are fixed, i.e. they are not varied
parametrically}. That means, the parametrical evolution of the 
different values, shown in the figures, occurs {\it without} changing
the coupling strength between system and environment.  

Let us consider  analytically the behavior that arises when the 
parametrical detuning of the two eigenstates of $\ch^{(2)}$ is varied, 
bringing them towards coalescence. 
According to (\ref{int6}), the  condition for coalescence reads 
\begin{eqnarray}
Z = \frac{1}{2} \sqrt{(e_1-e_2)^2 - \frac{1}{4} (\gamma_1-\gamma_2)^2 
+i(e_1-e_2)(\gamma_1-\gamma_2) + 4\omega^2} ~=~ 0 \; .
\label{int6i}
\end{eqnarray}
When $\gamma_1 = \gamma_2$, and  $\omega = i\,\omega_0$ is 
{\it imaginary}, it follows from (\ref{int6i})  
\begin{eqnarray}
(e_1 - e_2)^2 -4\, \omega_0^2 &= &0 
~~\rightarrow ~~e_1 - e_2 =\pm \, 2\, \omega_0 
\label{int6b}
\end{eqnarray}
such that two EPs appear. It furthermore holds 
\begin{eqnarray}
\label{int6c}
(e_1 - e_2)^2 >4\, \omega_0^2 &\rightarrow& ~Z ~\in ~\Re \\
\label{int6d}
(e_1 - e_2)^2 <4\, \omega_0^2 &\rightarrow&  ~Z ~\in ~\Im 
\end{eqnarray}
independent of the parameter dependence of $e_{1,2}$. According to
these equations, width bifurcation of states which are nearby in energy, 
causes the {\it formation of different time scales} in the system.  
The corresponding numerical results are shown in Fig. \ref{fig1} left panel. 
At the two EPs, $r_i \to 0$ and $|b_{ij}| \to \infty$. Between 
the two EPs, the widths bifurcate. In approaching the maximum width 
bifurcation (at the crossing point $e_1=e_2$), 
the phase rigidity of the states approaches rapidly the value $r_i \to
1$. That means, the wavefunctions become (almost) orthogonal at
this point. This result is completely unexpected.
    
An analog situation occurs in the case of Fig. \ref{fig2} left
panel. Here, $e_1 = e_2$ and $\omega $ is {\it real}. 
Instead of (\ref{int6b}) to (\ref{int6d}) we have
\begin{eqnarray}
(\gamma_1 - \gamma_2)^2 -16\, \omega^2 &= &0 
~~\rightarrow ~~\gamma_1 - \gamma_2 =\pm \, 4\, \omega 
\label{int6e}
\end{eqnarray}
and
\begin{eqnarray}
\label{int6f}
(\gamma_1 - \gamma_2)^2 >16\, \omega^2 &\rightarrow& ~Z ~\in ~\Im \\
\label{int6g}
(\gamma_1 - \gamma_2)^2 <16\, \omega^2 &\rightarrow&  ~Z ~\in ~\Re \; .
\end{eqnarray}
In this case,  states with comparable lifetimes {\it repel each other  
in energy}, see Fig. \ref{fig2} left panel. Like in Fig. \ref{fig1} left,
$r_i \to 0$ and $|b_{ij}| \to \infty$ at the two EPs. Although the 
levels repel each other in energy between the two EPs (while their 
widths are equal), the phase rigidity of the states approaches rapidly
the value $r_i \to 1$ also in this case at a critical parameter value 
between the two EPs. Here, level repulsion is maximum; and the 
wavefunctions are (almost) orthogonal. Also this result is unexpected. 
 
The figures in the right panels of Figs. \ref{fig1} and \ref{fig2}
show numerical results for the more realistic cases with complex
coupling coefficients $\omega$. They show the same characteristic
features as those in the corresponding left panels. In any case, most 
interesting is the parameter range between the position of an EP and that of, 
respectively,  maximum width bifurcation and maximum level repulsion. 
In this parameter range, the parametrical evolution of the system is driven 
{\it exclusively} by the nonlinear source term of the Schr\"odinger equation 
(\ref{form3a}) since -- as mentioned above --
the coupling strength $\omega$ between system and environment 
is fixed in the calculations. The analytical results, (\ref{int6b}) to 
(\ref{int6g}), support this statement. 

In Fig. \ref{fig1}, we have shown the case of
imaginary and almost imaginary coupling strength $\omega$ for the case
of decaying states ($\gamma_i < 0$ for $i=1,2$) and  
in Fig. \ref{fig2} the case with real and almost real $\omega$ for the case
with loss and gain ($\gamma_1 <0,~\gamma_2 > 0$). The
analytical and numerical results are  independent of the
relation between, respectively,  real and imaginary $\omega$ on the one hand,
and the possibility whether or not gain will take place, on the other hand.

\section{Discussion of the results}
\label{disc}

The results of our calculations shown in Sect. \ref{num}
are performed by keeping fixed the coupling strength
$\omega$ between system and environment. Width bifurcation (Fig. \ref{fig1})
and level repulsion (Fig. \ref{fig2}) can occur therefore only under the
influence of the source term of the Schr\"odinger equation, which is nonlinear
in the neighborhood of an EP, see (\ref{form3a}). According to the
results for the corresponding eigenfunctions, most interesting is the 
parameter range between the EP at which width bifurcation (level 
repulsion) starts, and that parameter value at which it becomes maximum. 
Here, the phase rigidity varies rapidly between its minimum and maximum
values. In more detail, it vanishes at an EP
according to the expectation (see section \ref{eig}) and increases, 
completely unexpected, up to its maximum value 
(nearly one) when width bifurcation (level repulsion) is maximum.  
According to the definition (\ref{eif11}) of the phase rigidity, this 
means that the two eigenfunctions are (almost) orthogonal when width
bifurcation (level repulsion) is maximum. Here, the eigenfunctions are mixed
in the set of the original wavefunctions, what can be seen best in the 
results shown in the left panels of Figs. \ref{fig1} and \ref{fig2}.  

The physical meaning of this result consists in the fact that
the localized part of the system gets stabilized. 
In the case of Fig. \ref{fig1}, one of the two
states receives a very short lifetime due to  width bifurcation, 
and becomes (almost) indistinguishable
from the  states of the environment. Although this process seems
to be reversible according to the subfigures for the {\it eigenvalues}, 
this is in reality not the case as can be seen from the subfigures 
for the {\it eigenfunctions}. As mentioned above, the processes 
occurring in approaching $r_k\to 1$, take place
exclusively  by means of the nonlinear source term of the Schr\"odinger 
equation (\ref{form3a}) near to an EP (since $\omega$ is fixed in the 
calculations). They provide states that are non-analytically connected 
to the original states.  Due to these 
processes, the long-lived state has ``lost'' its short-lived   
partner with the consequence that the two original states cannot be 
reproduced by further variations of the parameter. The parametric 
evolution is irreversible. The long-lived state is more stable than the 
original one (according to its longer lifetime), and the system
as a whole (which has lost one state)  is more stable than originally. 
The wavefunction of this long-lived state is mixed in those of the
original states.

In the case of Fig. \ref{fig2}, level repulsion of states with 
similar lifetimes causes a  separation of the states in energy.   
Due to level repulsion, the two states separate from one another 
in energy, such that their interaction
with one another is, eventually, of the same type as that with all 
the other distant states of the system. That means,  each of the original
states has ``lost'' its partner, also in this case; and the reproduction of  
the two originally  neighbored states is prevented. As a result, 
the interaction of the states of the system via the environment 
is reduced (since all states are distant in energy). 
In difference to the case with width bifurcation, however, the number
of states of the system as a whole remains unchanged.

According to this discussion, the results shown in Figs. \ref{fig1} 
and \ref{fig2} can be understood since the parametrical evolution
of the system is irreversible (what is marked in the figures by the 
change of the color of all the trajectories at the final point of 
respective evolution).
Eventually, the system as a whole is stabilized at the point of, 
respectively, maximum width bifurcation and maximum level repulsion;
can be described quite well by a Hermitian Hamilton operator the
eigenfunctions of which are orthogonal (corresponding to $r_i \approx 1$);
and the finite lifetime of the states is hidden, to a great extent.

Thus, our results presented in Figs. \ref{fig1} 
and \ref{fig2} show the important role, which the singular 
EPs play in open quantum systems when the system consists of a number
of {\it neighboring,  typically individual states}. 
The parametrical evolution of a system with more than two states is
studied in \cite{cluster}.  The irreversible processes of
stabilization of the system appear also in this case. They are
even stronger in the multi-level case  than in the two-level
case considered here. They cause, in an open quantum system,
some clustering of EPs; and dynamical phase transitions 
occurring in a critical parameter range.

\section{Relation to the phenomenon of resonance trapping}
\label{trap}

\begin{figure}[ht]
\begin{center}
\includegraphics[width=8cm,height=13cm]{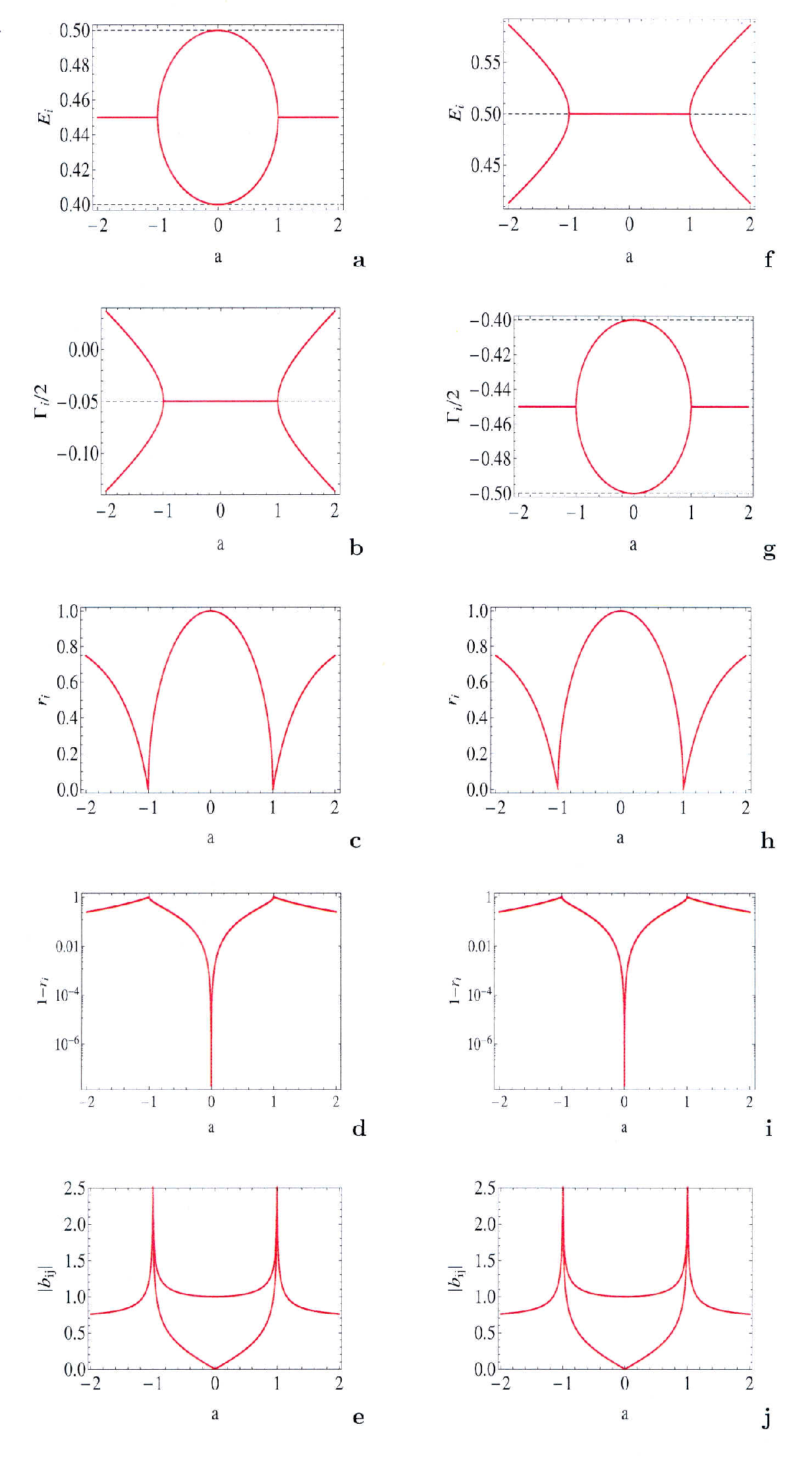}
\vspace{-.6cm}
\end{center}
\caption{\footnotesize
The same as Fig. \ref{fig1} but $\omega = \omega(a)$ and
fixed energies $e_i$ and $\gamma_i$.\\
Parameters:
$e_1=0.5; ~ e_2 = 0.4$ and $\gamma_1 = \gamma_2 = -0.05$; 
$\omega = 0.05\,a\;i$ (left); 
$e_1 = e_2=0.5$ and $\gamma_1 =-0.5; \gamma_2=-0.4 $; 
$\omega =0.05\,a $ (right).\\
The dashed lines in (a, b, f, g) show $e_i =$ const
and $\gamma_i =$ const, respectively.
The results for the phase rigidity $r_i$ 
are analog to those in Figs. \ref{fig1} and \ref{fig2},
while those for the mixing coefficients $|b_{ij}|$ are different.
We used the same color  in the whole parameter range 
for the different trajectories in order to express the reversible
parametric evolution when $\alpha $ depends linearly on the
parameter $a$   (see the  discussion in the text). 
}
\label{fig3}
\end{figure}

The results shown in section \ref{num} and discussed
in section \ref{disc} should not be confused with the
phenomenon of resonance trapping \cite{ro91}, called also segregation 
of resonance widths or superradiance \cite{auerbach}. This effect is 
known for more than 15 years and discussed controversially in literature 
still today. It is nothing but width bifurcation
occurring in an open quantum system with many {\it randomly
distributed levels}, which is described by a non-Hermitian Hamiltonian.
The effect appears in a large parameter range when the coupling strength 
$\omega$ between system and environment is parametrically enhanced.
{\it It is difficult therefore to relate it to a phase transition}. 
Mostly, the approximation $\ch = H^B - i \alpha VV^+$ is used for the
description of the system where $H^B$ is a Hermitian operator; 
$VV^+$ stands for the non-Hermitian perturbation; and $\alpha$ 
is the tuning parameter \cite{ro91,top,auerbach}.
Eventually,  long-lived trapped states are separated from the 
short-lived state (in the one-channel case), see e.g. 
\cite{harney,jumuro}.
The influence of EPs is not considered in these papers since
individual states are typically not looked at in these papers. 

For illustration, we show in Fig. \ref{fig3} numerical results for
the case that {\it $\omega$ is parametrically varied while the energies
$e_i$  and widths $\gamma_i$ are fixed}. The condition (\ref{int6i})
for $Z=0$ has to be fulfilled also in this case. This causes 
results which are similar to those considered 
in detail in Eqs. (\ref{int6b}) to (\ref{int6g}) and 
in Figs. \ref{fig1} and \ref{fig2}. Indeed, the phase 
rigidity approaches  the value zero at an EP while it approaches 
the value one at the point of maximum width bifurcation and maximum
level repulsion, respectively. The shape of the $r_i$ trajectories 
at the EPs is determined by the fact that the variation of $r_i$ 
occurs more slowly at one side than at the other one.

The wavefunctions of the
states show, at first glance,  another behavior 
than those in Figs. \ref{fig1} and \ref{fig2}. The tuning
parameter $\alpha$  (which characterizes the {\it mean} coupling 
strength of the states of the system to the environment)
is assumed, usually, to be real and to increase linearly from  
zero up to $\infty$. At the critical value $\alpha = \alpha_{\rm cr}$, 
a restructuring of the system  takes place
\cite{jumuro,mudiisro}\,: all but one state become trapped (in the
one-channel case) while the width of one state increases further.  
Accordingly, the wavefunctions of the states are (almost) orthogonal 
at small $\alpha$ and become mixed beyond the restructuring (being a 
second-order phase transition) \cite{jumuro}. This picture arises
when -- in addition to the eigenvalues -- also the eigenfunctions 
are considered. It  can be seen qualitatively also in   
Fig. \ref{fig3}.e\,: the wavefunctions are {\it not} mixed  
when $\alpha $ approaches zero (and the levels are distant from 
one another), while they become mixed when  
$\alpha > \alpha_{\rm cr}$ (and the widths bifurcate);
see also Fig. 1 in \cite{mudiisro} where the convergence of
distant levels in approaching  $\alpha_{\rm cr}$ is shown.
In any case, the eigenfunctions are biorthogonal also for 
small $\alpha$ what is proven even experimentally \cite{savin}. 

The studies in section \ref{num} show that the nonlinear source term 
of the Schr\"odinger equation (\ref{form3a}) strengthens and accelerates 
(as a function of the parameter $a$) considerably the separation of 
the states around an EP in energy and lifetime, respectively. 
The appearance of a dynamical phase transition  will therefore be 
visible also in the representation of the results in a figure like
Fig. \ref{fig3} when the linear relation 
$\alpha \propto a$ for all $a$ (from $a=0$ to $\infty$)  
is replaced, at certain $a$ values, by a nonlinear one
due to the nonlinear source term of the Schr\"odinger equation
(\ref{form3a}) near an EP. In such a case, the phase 
rigidity will approach quickly the value $r_i \approx 1$ 
also for large $|a|$; and $\alpha$ will not increase limitless.
Correspondingly to this picture, 
the approximation $\ch = H^B - i \alpha VV^+$
can be applied to the description of a multi-level system {\it only}
near to and at the dynamical phase transition; and
the states on both sides of the transition are
non-analytically related to one another \cite{nearby2}.

In the paper \cite{auerbach}, the wavefunctions of the states are 
described randomly; and the approach that $\alpha$ depends 
everywhere linearly on $a$, is 
used. The results are related to the phenomenon of superradiance; and  
the approach is applied to the description of different processes in 
physical systems, above all in nuclear physics. In \cite{celardo} it 
is applied  to the energy transport in photosynthetic 
light-harvesting complexes.  The influence of
EPs is not considered in these papers. According to our  results 
and the discussion in Sects. \ref{num} and \ref{disc},   
superradiance (or resonance trapping)  is a clear hint to a 
dynamical phase transition.

\section{Concluding remarks}
\label{concl}

The basic concept of our formalism is supported by
many experimental results received in different physical systems, 
mostly in mesoscopic systems, see the review \cite{ropp}. 
The relation between system and environment on the one
hand and the measuring process on the other hand is
illustrated for a mesoscopic system in  Fig. 1 of \cite{ropp}.

Another experimental result concerns the 
tunneling effect due to which, according to standard quantum mechanics, 
electrons can escape from atoms under the influence of strong laser 
fields. Recently, the tunneling delay time is measured in Helium 
by attosecond ionization \cite{eckle}. The experimental results 
{\it give a strong evidence that there is no real tunneling 
delay time} \cite{eckle}.
This result is in complete agreement with (\ref{int6}) according to 
which the escape time is  nothing but the imaginary part of the 
eigenvalue of the non-Hermitian Hamiltonian,  
$\Gamma_i \propto {\rm Im}(\ce_{i})$. 

Whether or not the new experimental results \cite{hanson}
on the violation of the Bell inequality can be related to the 
basic assumptions of our calculations needs further investigations. 
See also the papers \cite{merali,wiseman}. In any case,  
a 'spooky action at a distance' is possible only at and near to an 
EP where the interaction via the environment between two states 
approaches an infintely large value, as the results of our 
calculations show (Figs. \ref{fig1} and \ref{fig2}).

As a conclusion we state that there are some differences between a
dynamical  phase transition in an open quantum system and the
phenomenon of PT-symmetry breaking, in spite of many obvious
similarities. The dynamical phase transition is a very
general and robust phenomenon. We expect   interesting 
results similar to those obtained for PT-symmetric systems 
in optics and photonics with balanced loss and gain,  
in the near future also in other systems.


\begin{thebibliography}{99}
\bibitem{comment1}  The coalescence of two eigenvalues of a
  non-Hermitian operator should not be confused with the degeneration
  of two eigenstates of a Hermitian operator. The eigenfunctions of two 
degenerate states are  different and orthogonal while those of two
coalescing states are biorthogonal and differ only by a phase, 
see Eqs. (\ref{int3}) to (\ref{eif5}).
\end{thebibliography}
\end{document}